\documentclass[twocolumn]{aastex631}
\let\tablenum\relax

\usepackage{placeins}
\usepackage{siunitx}

\shorttitle{$r$-Process Abundances in Globular Clusters}
\shortauthors{Kenoly et al}

\graphicspath{{./}{figures/}}

\begin{document}

\correspondingauthor{\\
$^{\dagger}$Lailani Kenoly: \href{mailto:lkenoly21@gmail.com}{lkenoly21@gmail.com} \\
$^{\ddagger}$Isabelle Connor: \href{mailto:iconnor@princeton.edu}{iconnor@princeton.edu}}

\title{$r$-Process Abundance Variations in Globular Clusters as Fossils of Turbulent Mixing}

\author{Lailani Kenoly$^\dagger$}
\affiliation{Department of Physics \& Astronomy, San Jose State University, San Jose, CA 95192, USA}
\affiliation{Department of Astronomy \& Astrophysics, University of California, Santa Cruz, CA 95064, USA}

\author[0009-0008-3986-657X]{Isabelle  Connor$^\ddagger$}

\affiliation{Department of Astrophysical Sciences, Nassau Street, Princeton University, Princeton, NJ 08544, USA}

\author[0000-0001-7493-7419]{Melinda Soares-Furtado}
\affiliation{Department of Astronomy, University of Wisconsin-Madison, 475 N. Charter Street, Madison, WI 53706, USA} 
\affiliation{Department of Physics, 2320 Chamberlin Hall, University of Wisconsin-Madison, 1150 University Avenue, Madison, WI 53706}

\author{Evan N. Kirby}
\affiliation{Department of Physics, University of Notre Dame, Notre Dame, IN 46556, USA}

\author{Nia Imara}
\affiliation{Department of Astronomy \& Astrophysics, University of California, Santa Cruz, CA 95064, USA}

\author[0000-0003-2558-3102]{Enrico Ramirez-Ruiz}
\affiliation{Department of Astronomy \& Astrophysics, University of California, Santa Cruz, CA 95064, USA}

\begin{abstract}
Several metal-poor Galactic globular clusters exhibit substantial
star-to-star $r$-process abundance variations at nearly fixed [Fe/H], with
the ensemble suggesting smaller variations toward increasing metallicity.
We investigate whether this chemical structure can be inherited from
incomplete mixing in the galactic environments in which the clusters formed.
Using parsec-scale simulations of rare-element transport in a
supernova-driven turbulent disk, we follow the redistribution of $r$-process
ejecta and characterize the abundance structure of gas selected to trace
star-forming environments. Rare enrichment events generate strong [Eu/Fe]
inhomogeneities that persist across a range of physical scales. As chemical evolution proceeds, the available abundance variance decreases, while spatial averaging over larger regions reduces the differences between
their characteristic abundances. The simulations provide substantial pre-collapse
$r$-process variance at low [Fe/H] and show the same broad decline toward
higher metallicity suggested by the globular-cluster measurements. Our
results support an interpretation in which at least some of the observed
$r$-process variations are inherited from pre-existing chemical structure
in the natal environment, allowing globular clusters to preserve a fossil
record of the mixing history between rare enrichment and star formation.
\end{abstract}


\section{Introduction}
\label{sec:intro}

Globular clusters (GCs) preserve detailed records of the chemical environments
in which they formed. While most are remarkably homogeneous in iron, several
metal-poor clusters show significant star-to-star variations in neutron-capture
elements at nearly fixed [Fe/H]. \citet{Roe2011} showed that these
variations are dominated by the $r$-process and are largely decoupled from the
light-element abundance patterns associated with multiple stellar populations.
Subsequent observations have confirmed substantial $r$-process dispersion in
M15, M92, and other metal-poor clusters, while other systems show much smaller
or undetectable variations
\citep{2015MNRAS.449.3889R,2017ApJ...842...24J,
2023ApJ...958...45K,2024ApJ...967..101C,2025AJ....170...37B,
2025ApJ...990..132N,2025ApJ...992L..14H}. Taken together, these measurements
suggest that the amplitude of the internal $r$-process variation is largest
among some of the most metal-poor clusters and generally decreases toward
increasing metallicity.

The origin of these variations remains uncertain. M15 provides an important
constraint: the absence of dilution of its Ba dispersion with stellar evolution
demonstrates that the stars inherited their heavy-element abundances at birth
\citep{2020ApJ...891L..13K}. Moreover, recent observations of M15 and M92 find
larger $r$-process dispersion among first-generation stars than among later
populations \citep{2023ApJ...958...45K,2025ApJ...992L..14H}.
These results favor a picture in which the abundance fluctuations were already
present in the gas from which the stars formed, with subsequent mixing reducing
the chemical variance inherited by later-forming populations
\citep[e.g.,][]{2015ApJ...807..115S,2018MNRAS.477.1206N}.

Several models have instead considered enrichment by sources associated with
the forming cluster, including compact-object mergers whose ejecta pollute gas
retained or reaccreted by the system
\citep[e.g.,][]{2010ApJ...720..953L,2015ApJ...802L..22R,2019ApJ...886....4Z}.
Alternatively, the chemical structure may predate the cluster itself.
\citet{2021ApJ...921L..11T} showed that inhomogeneous enrichment of the natal
environment before cluster formation can produce an Eu distribution resembling
that of M15. This possibility suggests a broader interpretation: GCs may
preserve fossil signatures of incomplete mixing in the galactic environments
from which they assembled.

Here we test this idea using parsec-scale simulations of rare-element transport
in a supernova-driven turbulent galactic disk
\citep{2022ApJ...936L..26K,2023ApJ...949..100K,2025ApJ...994..193B}.
Rare enrichment events initially generate spatially inhomogeneous
$r$-process abundance fields that are subsequently redistributed by turbulent
motions. Rather than modeling GC formation itself, we characterize the
pre-stellar chemical variance in the surrounding interstellar medium and ask
how it evolves with chemical enrichment and physical scale. In particular, we
distinguish between abundance variations among different spatial regions and
the chemical variance contained within an individual region, the latter being
more directly relevant to the star-to-star variations ultimately preserved
within a cluster. We ask whether the observed diversity and metallicity
dependence of internal $r$-process variations in GCs can be understood as the
surviving imprint of pre-existing chemical structure shaped by turbulent
transport.

Our treatment is agnostic to the identity of the $r$-process progenitor
\citep[e.g.,][]{2017Natur.551...80K,2019EPJA...55..203S} once its ejecta are
deposited into the galactic disk: the enrichment event may represent a rare
supernova or a compact-object merger
\citep[e.g.,][]{2018ApJ...860...89M,2019ApJ...872..105S,
2019ApJ...877L..24M,2022NatRP...4..306S}. We therefore isolate the transport
problem from the progenitor problem. The latter determines whether an event
occurs in the relevant star-forming environment, an important consideration
for mergers with long delay times or significant systemic velocities, whereas
our simulations determine how the resulting chemical structure evolves once
the material is injected \citep{2023ApJ...949..100K}. In this framework, the
internal $r$-process variations of GCs provide a fossil record of the
competition between stochastic enrichment and pre-stellar turbulent mixing.
\section{Modeling Chemical Mixing in a High-SFR Milky Way Progenitor}
\label{sec:methods}

The distribution of $r$-process material in galaxies reflects the competition
between rare, stochastic enrichment events and turbulent transport in the
ISM. We use the high star formation rate (high-SFR) simulation of
\citet{2023ApJ...949..100K}, which reproduces both the mean $r$-process
abundance and the broad abundance dispersion observed among metal-poor
Milky Way stars. Here we use the same calculation to address a different
question: how is the $r$-process abundance field structured on finite spatial
scales, both between distinct regions and within the regions from which
stellar systems such as globular clusters may subsequently assemble?

The simulations were performed with the adaptive-mesh hydrodynamics code
\textsc{RAMSES} \citep{2002A&A...385..337T} and follow a three-dimensional,
vertically stratified patch of a galactic disk
\citep{2016MNRAS.459.2311M,2022ApJ...936L..26K,2023ApJ...949..100K}.
The high-SFR calculation follows a $1$ kpc box for $120$ Myr with a spatial
resolution of $\Delta x=3.9$ pc, corresponding to $256^3$ cells. Periodic
boundary conditions are imposed in the disk plane and outflow conditions
perpendicular to it. The gaseous disk has an effective scale height of
$\approx40$ pc.

Core-collapse supernovae (CCSNe) drive the turbulent ISM, while radiative
cooling and an imposed gravitational potential produce a multiphase,
vertically stratified gas distribution. Metals are advected with the gas,
allowing their spatial redistribution to be followed directly. The cooling
radius of individual supernova remnants is resolved by at least five cells
for more than 94\% of the events in this model
\citep{2022ApJ...936L..26K,2023ApJ...949..100K}. Baroclinic vorticity
generated at the unstable cooling interfaces of expanding supernova remnants,
together with interactions between remnants, drives turbulent motions over a
broad range of scales, including non-local transfer between the injection and
larger disk scales
\citep{2025ApJ...994..193B,2026ApJ...997...33C,2026ApJ..1004L...9B}.
The calculation therefore captures the supernova-driven turbulent transport
that redistributes freshly injected material, without imposing an explicit
subgrid metal-mixing prescription.

The high-SFR model corresponds to an implied star formation rate of
$\approx10\,M_\odot\,{\rm yr}^{-1}$ and a CCSN rate of
$\approx0.1\,{\rm yr}^{-1}$. CCSNe are seeded randomly at a constant rate,
with a flat spatial distribution within a maximum height of $100$ pc from
the disk midplane \citep{2023ApJ...949..100K}. Iron is injected by the
frequent CCSNe, whereas $r$-process material is introduced through rare
stochastic events with a rate $f_{\rm rp}=10^{-2}$ relative to the CCSN rate
and an ejecta mass $m_{\rm rp}=10^{-2}\,M_\odot$ per event
\citep{2023ApJ...949..100K}. These parameters correspond to an average
$r$-process production rate of
$\approx10^{-5}\,M_\odot\,{\rm yr}^{-1}$. Eu is used as a tracer of the
$r$-process material \citep{2000ApJ...544..302B}. The injection prescription
does not require us to specify the astrophysical identity of the rare event:
once its ejecta are deposited into the disk, we follow their subsequent
transport and mixing irrespective of whether the source is a rare supernova
or a compact-object merger.

\begin{figure}
    \centering
    \includegraphics[width=0.47\textwidth]{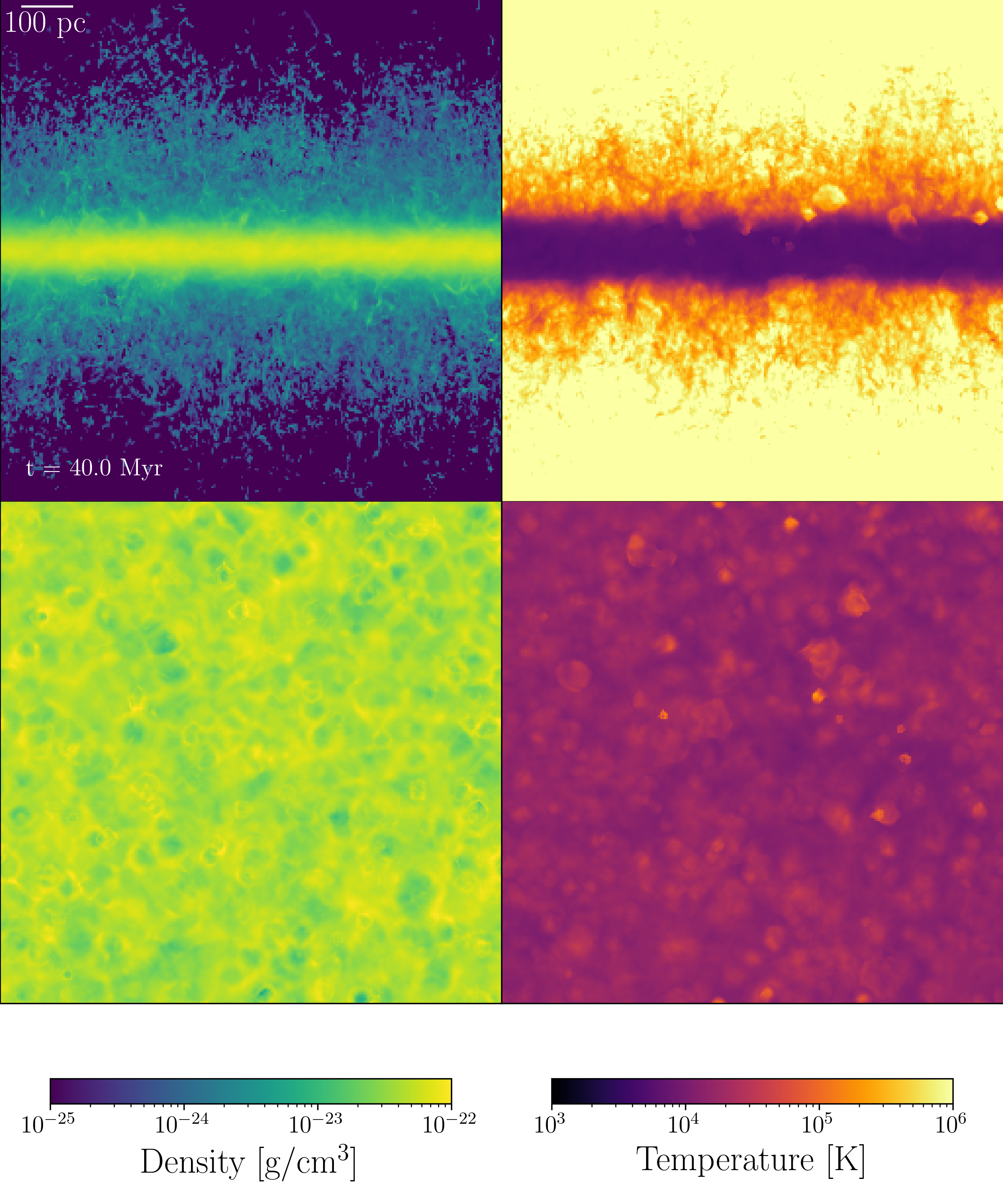}
    \caption{Density-weighted projections of the gas density (left) and
    temperature (right), viewed perpendicular to the disk plane (top) and
    parallel to it (bottom). Individual supernova remnants appear as bubbles
    of low-density, high-temperature gas, illustrating the multiphase,
    supernova-driven environment through which freshly synthesized material
    is transported.}
    \label{fig:rhoT_maps}
\end{figure}

The simulations are not intended to model the formation of individual
globular clusters. In particular, they do not resolve the detailed
thermochemistry of the cold molecular ISM, gravitational collapse,
fragmentation, or the subsequent formation of a stellar cluster. Gas with
$T\leq10^4$ K should therefore be regarded as tracing the cooler component
represented by the calculation rather than as a resolved model of molecular
gas. Nevertheless, the simulations capture the larger-scale supernova-driven
transport that establishes the chemical environment from which colder
star-forming structures may subsequently emerge \citep{2016MNRAS.459.2311M,2020ApJ...896...66K,
2025ApJ...994..193B,2026ApJ...997...33C,2026ApJ..1004L...9B}.

Turbulent transport in
these calculations is regulated by vorticity generated at radiative cooling
interfaces and by interactions between expanding remnants
\citep{2025ApJ...994..193B,2026ApJ...997...33C,2026ApJ..1004L...9B}.
Our analysis therefore characterizes the pre-collapse chemical abundance
field from which globular clusters may subsequently assemble, rather than
predicting their final stellar abundance distributions directly.

Figure~\ref{fig:rhoT_maps} illustrates this thermodynamic structure.
Repeated supernova explosions generate a highly structured, multiphase
medium, with hot, low-density remnants embedded within cooler, denser gas
concentrated toward the disk.

To characterize the abundance field on finite spatial scales, we identify
gas with $T\leq10^4\,{\rm K}$ and
$\rho\geq10^{-23}\,{\rm g\,cm^{-3}}$ and randomly select 100 cells satisfying
these criteria at each simulation output. These cuts select the cooler,
denser component represented by the calculation and should not be interpreted
as selecting resolved molecular gas. Around each selected cell we construct
spherical sampling regions of radius $R$, spanning $R\approx8$--$120$ pc.
These regions provide a controlled way of measuring how the chemical
structure of the simulated gas changes with spatial scale. The smallest
regions span only a few grid cells and should therefore be interpreted
cautiously, since finite spatial resolution and numerical diffusion
increasingly affect the abundance structure on these scales.

For each region $j$ of radius $R$, we characterize its abundance by the
median
\begin{equation}
    A_j(R)
    =
    {\rm median}_j\left([\mathrm{Eu}/\mathrm{Fe}]\right),
\end{equation}
and measure its internal abundance variation using the standard deviation
\begin{equation}
    \sigma_j(R)
    =
    \left[
    \frac{1}{N_j}
    \sum_{i=1}^{N_j}
    \left(
    A_{ij}-\overline{A}_j
    \right)^2
    \right]^{1/2},
\end{equation}
where $A_{ij}=[\mathrm{Eu}/\mathrm{Fe}]_{ij}$, $N_j$ is the number of cells
within the region, and $\overline{A}_j$ is their mean abundance.

These quantities characterize two complementary aspects of the chemical
field. The distribution of $A_j(R)$ describes how the characteristic
abundance varies among spatially distinct regions, while $\sigma_j(R)$
measures the abundance variation contained within an individual region.
Examining both quantities over a range of $R$ allows us to distinguish
chemical variations between regions from those that persist within them.
Together they provide a scale-dependent description of the pre-collapse
abundance field from which stellar systems may subsequently assemble.

\begin{figure*}
    \centering
    \includegraphics[width=0.8\textwidth]{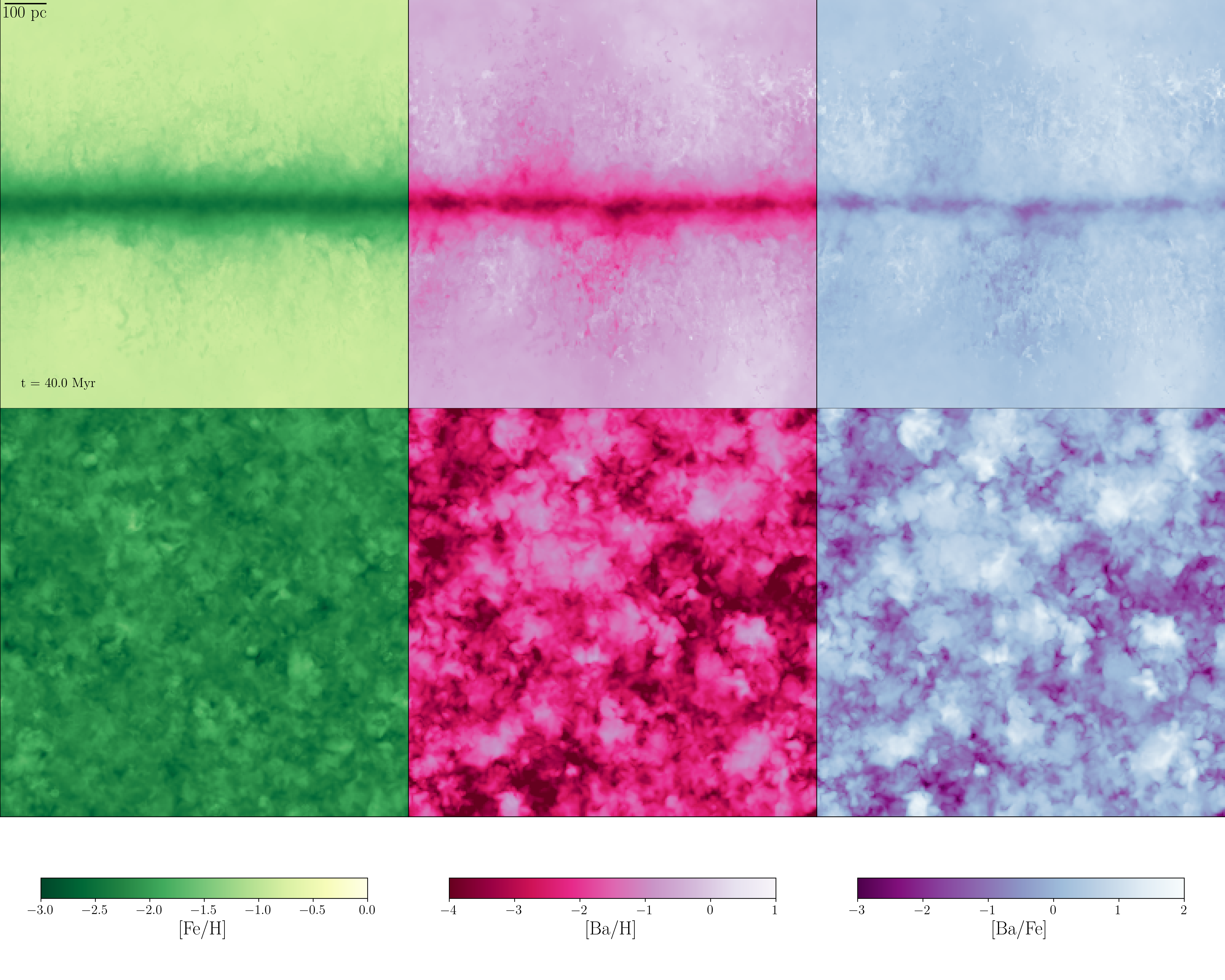}
    \caption{Density-weighted projections of the gas abundances
    (from left to right): [Fe/H], [Ba/H], and [Ba/Fe], viewed perpendicular
    to the disk plane (top) and parallel to the disk (bottom). Ba is shown
    here as an $r$-process tracer, motivated by its use in globular-cluster
    abundance studies such as M15 \citep{2020ApJ...891L..13K}; the
    corresponding Eu abundance field from these simulations is shown in
    \citet{2023ApJ...949..100K}. Ba and Eu trace the same underlying
    $r$-process enrichment field, whose ejecta are redistributed by the
    resolved hydrodynamics down to the resolution scale of the simulation.
    We use Eu for the quantitative analysis below.}
    \label{fig:abundance_maps}
\end{figure*}

\section{The Chemical Imprint of Rare Enrichment}
\label{sec:chemical_imprint}

The rarity of $r$-process enrichment produces a chemical abundance field that
is qualitatively different from that of elements supplied by frequent
core-collapse supernovae. Figure~\ref{fig:abundance_maps} shows the spatial
distribution of [Fe/H], [Ba/H], and [Ba/Fe]. By this stage of the simulation,
repeated CCSNe have produced a comparatively smooth Fe distribution throughout
the disk. In contrast, Ba retains pronounced spatial structure, with localized
regions of strong enrichment surrounded by gas with substantially lower
$r$-process abundance. The resulting [Ba/Fe] field therefore preserves the
imprint of rare enrichment events even after their ejecta have been
transported through the turbulent ISM. This difference reflects the sampling
of the two enrichment channels: many CCSNe contribute to the local Fe
inventory, whereas the $r$-process abundance can remain dominated by only a
small number of rare events
\citep{2022ApJ...936L..26K,2023ApJ...949..100K}.

\begin{figure}
    \centering
    \includegraphics[width = 0.45\textwidth]{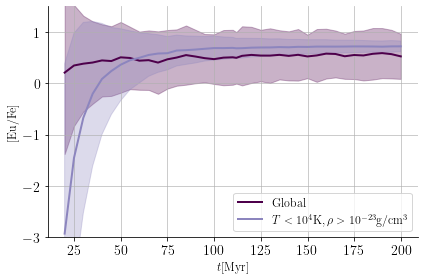}
    \caption{The mean and $1\sigma$ spread of the gas
    $[\mathrm{Eu}/\mathrm{Fe}]$ abundance as a function of time. The dark
    purple curve includes all gas in the simulation volume, while the light
    purple curve includes only gas with $T<10^4$ K and
    $\rho>10^{-23}\,{\rm g\,cm^{-3}}$. The mean abundances of the two
    selections converge as enrichment proceeds, while their dispersions
    evolve differently. The full gas volume retains substantial abundance
    fluctuations, whereas the $r$-process dispersion in the selected gas
    decreases progressively with time.}
    \label{fig:EuFe_evolution}
\end{figure}

The abundance fluctuations contain information that is not captured by the
mean abundance alone. Figure~\ref{fig:EuFe_evolution} follows the mean
[Eu/Fe] abundance and its $1\sigma$ spread in the full simulation volume and
in the gas selected by the temperature and density criteria described in
Section~\ref{sec:methods}. This statistic characterizes the abundance
distribution over each gas selection as a whole and is distinct from the
finite-region measurements introduced in Section~\ref{sec:methods}. The
global mean rapidly approaches
$[\mathrm{Eu}/\mathrm{Fe}]\approx0.25$ and evolves only slowly thereafter.
The full simulation volume nevertheless maintains a broad abundance
distribution, with a dispersion of order $0.5$ dex over much of the
simulation. In contrast, the dispersion of the selected gas declines
steadily with time, even as its mean abundance approaches that of the full
volume.

This distinction is important for connecting galactic chemical structure to
the environments from which stellar systems form. The abundance distribution
of the ISM as a whole is not necessarily representative of the gas selected
to trace prospective star-forming environments. In the simulation, the
temperature- and density-selected gas becomes progressively more chemically
homogeneous as enrichment and turbulent transport proceed, while substantial
$r$-process structure remains elsewhere in the volume. Globular clusters are
thought to form from dense, highly structured gas in high-pressure
star-forming environments
\citep[e.g.,][]{2015MNRAS.454.1658K,2019ARA&A..57..227K}.

The subsequent assembly and collapse of this gas introduce additional
opportunities for redistribution before its chemical structure is locked
into stars. Turbulent motions during cloud assembly can reduce chemical
inhomogeneities inherited by the stellar population
\citep{2014Natur.513..523F,2018MNRAS.481.5000A}. Once massive stars begin to
form, stellar winds provide additional energy and momentum to the surrounding
gas \citep[e.g.,][]{2014MNRAS.442.2701R} and can drive turbulence within
young massive clusters \citep[e.g.,][]{2020ApJ...899L..30G}. These processes
are not included in our calculation and may further redistribute material
during the unresolved stages of cloud assembly, collapse, fragmentation, and
star formation. The abundance structure measured here should therefore be
interpreted as the pre-collapse chemical field available to a
cluster-forming environment, rather than as a direct prediction of the final
star-to-star abundance dispersion.

The stellar fossil record should consequently reflect the chemical structure
that survives these subsequent stages of cluster formation rather than the
abundance distribution of the ISM as a whole. As the simulation evolves, the
$r$-process dispersion in the selected gas declines while its mean abundance
continues to increase. This establishes the central trend that we will use to
connect the evolving galactic abundance field to the metallicity dependence
of $r$-process variations observed in globular clusters.

The dispersion across the selected gas as a whole, however, does not specify
how that chemical structure is distributed on the finite spatial scales from
which an individual stellar system may assemble. We therefore next examine
how the characteristic abundance of sampled regions and the chemical
variation contained within them depend on the sampling scale $R$.

\section{The Scale Dependence of Chemical Memory}
\label{sec:scale_dependence}

The decline in the abundance dispersion of the selected gas shown in
Figure~\ref{fig:EuFe_evolution} describes its evolution when considered as a
whole. The abundance field, however, also contains spatial structure on a
range of physical scales. A large sampling region averages over more of this
structure, whereas a smaller region remains more sensitive to local abundance
variations. The sampling radius $R$ should therefore be understood as a
real-space averaging scale rather than as a turbulent scale associated with
a particular Fourier mode: each spherical region contains contributions from
a range of turbulent modes.

We quantify this spatial structure using the spherical sampling regions
defined in Section~\ref{sec:methods}.

\begin{figure}
\centering
\includegraphics[width=0.45\textwidth]{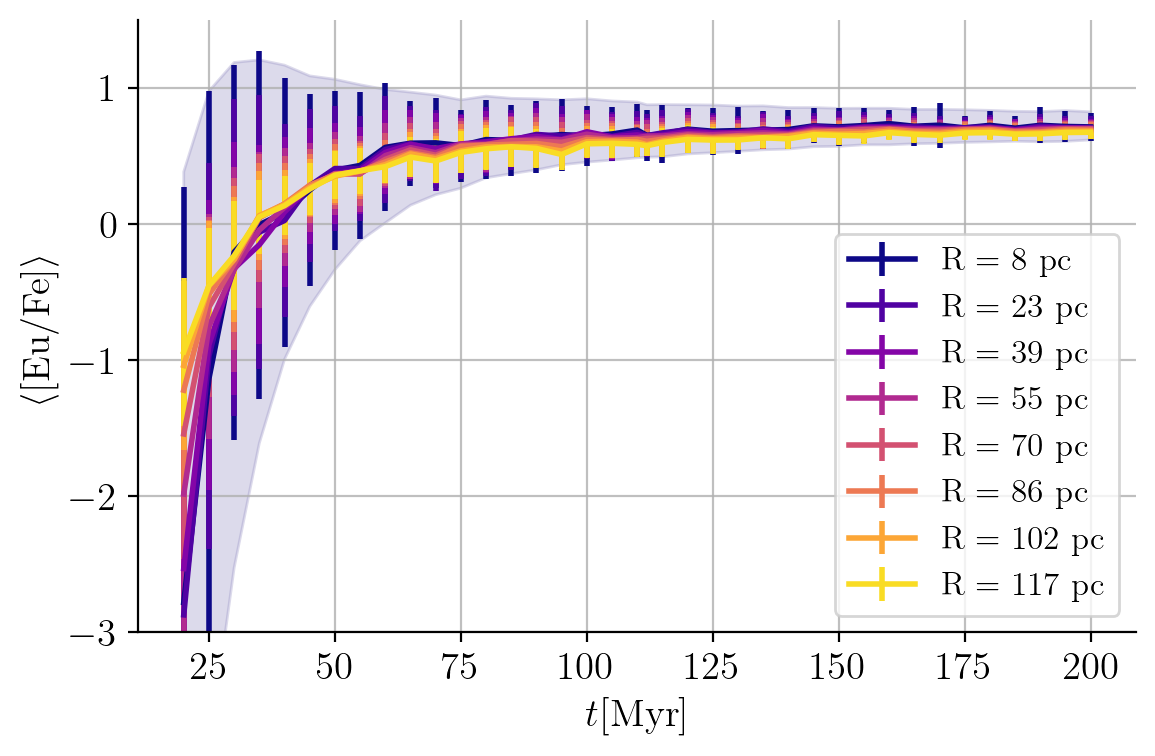}
\caption{Characteristic [Eu/Fe] abundance and region-to-region dispersion
for randomly selected spherical regions centered on gas satisfying the
temperature and density criteria defined in Section~\ref{sec:methods}, shown
as functions of time and sampling radius $R$. The ensemble abundance depends
only weakly on $R$, whereas the dispersion among regions increases toward
smaller sampling radii. Larger regions average over more of the spatial
abundance field, while smaller regions retain greater sensitivity to
localized chemical structure.}
\label{fig:EuFe_scale}
\end{figure}

Figure~\ref{fig:EuFe_scale} shows the characteristic [Eu/Fe] abundance
defined by the regional median, together with its region-to-region
dispersion, for regions with radii from $R\approx8$ to $120$ pc. The ensemble abundance
depends only weakly on sampling radius, whereas the dispersion among regions
increases systematically toward smaller $R$. Larger regions average over
more of the spatial abundance field and consequently approach similar
abundances. Smaller regions average over less of this structure and therefore
retain larger region-to-region abundance contrasts.

This scale dependence measures how the chemical structure of the gas appears
under spatial averaging. It should not by itself be interpreted as a direct
measurement of the turbulent cascade, since changing $R$ changes the volume
over which the abundance field is sampled rather than isolating fluctuations
at a particular turbulent wavenumber. Nevertheless, the spatial structure
being sampled is generated and redistributed by the supernova-driven
turbulence. Measurements of the velocity structure and temporal coherence in
these simulations identify characteristic scales of order tens of parsecs
for the turbulent flow
\citep{2025ApJ...994..193B,2026ApJ...997...33C,2026ApJ..1004L...9B}.
Our real-space sampling therefore asks how much of the chemical structure
generated and redistributed by this turbulence remains when the abundance
field is averaged over regions of different physical size.

The chemical memory of the gas therefore depends on both its evolution in
time and the physical scale over which it is sampled. At a given $R$, the
abundance structure evolves as enrichment and turbulent transport proceed.
At a given time, smaller regions retain larger region-to-region abundance
contrasts because they average over less of the underlying chemical
structure. The pre-collapse chemical environment available to a forming
stellar system is therefore determined by both the evolution of the abundance
field and the finite spatial region from which its material is assembled.

This distinction is particularly important for globular clusters. A cluster
does not sample the abundance field averaged over the full turbulent disk,
but instead assembles from a finite region embedded within this spatially
evolving chemical field. Its stellar abundances may therefore preserve
chemical structure present in its natal material even when that structure is
substantially reduced when averaged over larger spatial scales. We now ask
whether the observed $r$-process variations in Galactic globular clusters
are consistent with this picture.

\section{Globular Clusters as Fossils of Turbulent Mixing}
\label{sec:gc_fossils}

\begin{figure*}
    \centering
    \includegraphics[width=0.9\textwidth]{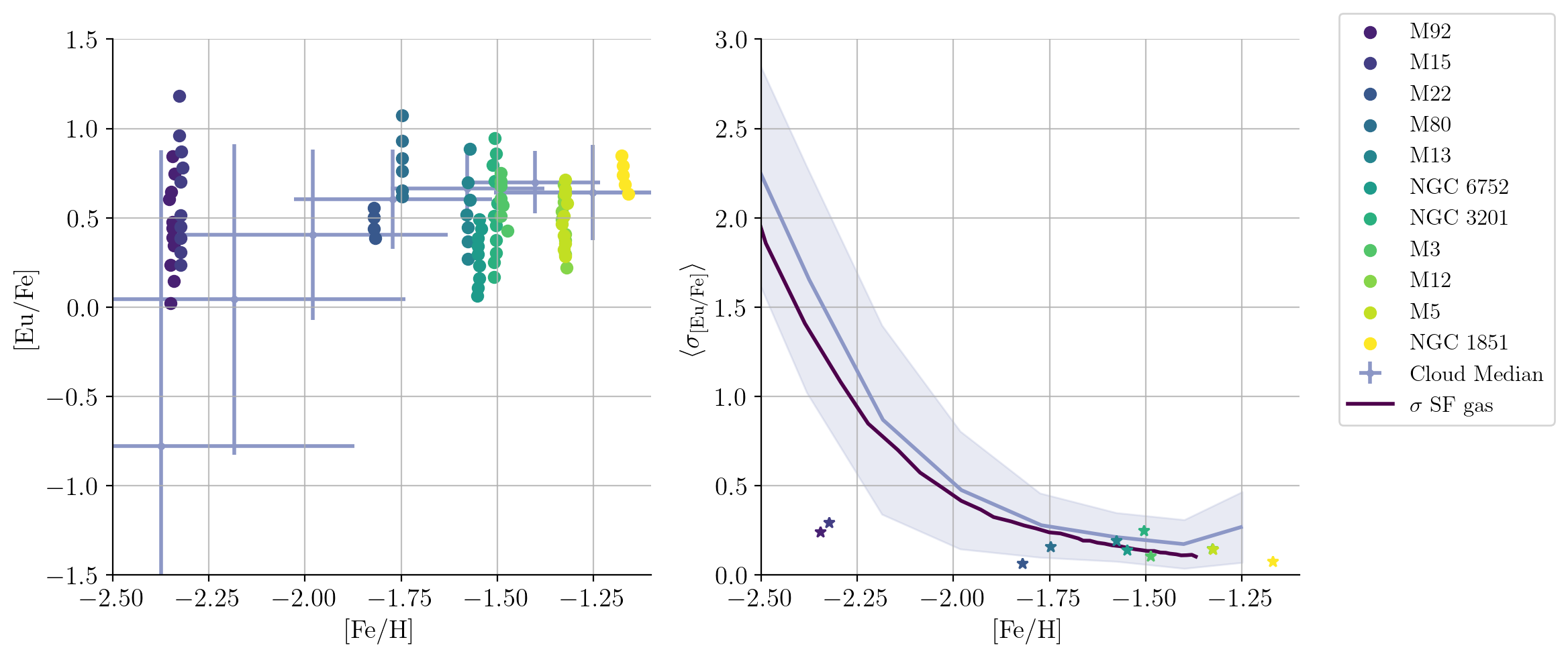}
    \caption{Comparison between the $r$-process abundance structure of
    Galactic globular clusters and the simulated gas.
    \textit{Left:} Stellar [Eu/Fe] measurements for individual globular
    clusters as a function of [Fe/H] \citep{Roe2011}. The simulation crosses show the mean
    of the median [Eu/Fe] abundances measured in sampled regions with
    $R=39$ pc; their vertical extent shows the cloud-to-cloud dispersion of
    these median abundances.
    \textit{Right:} Internal [Eu/Fe] dispersions of the observed clusters \citep{Roe2011}
    compared with two measures of the simulated abundance variance. The
    cloud-scale curve shows the mean internal standard deviation within the
    $R=39$ pc sampled regions, while the second curve shows the $1\sigma$
    dispersion of the selected gas considered as a whole. The larger
    box-wide dispersion demonstrates that substantial chemical variance
    remains available outside any individual sampled region, while the
    cloud-scale dispersion measures the variance more directly relevant to
    a prospective cluster-forming region. Both simulated measures decrease
    as chemical evolution proceeds.}
    \label{fig:obs}
\end{figure*}

Figure~\ref{fig:obs} provides the central observational comparison. To
minimize systematic differences between abundance analyses, we restrict the
primary comparison to globular clusters analyzed on a homogeneous abundance
scale \citep{Roe2011}. The left panel shows the measured [Eu/Fe] abundances of individual
stars in these clusters. Although each cluster spans only a narrow range in
[Fe/H], several of the most metal-poor systems exhibit substantial
star-to-star $r$-process variations. The largest variations occur at low
metallicity, while the overall amplitude tends to decrease toward increasing
[Fe/H]
\citep{Roe2011,2020ApJ...891L..13K,
2023ApJ...958...45K,2024ApJ...967..101C,2025ApJ...992L..14H}.
The evidence is not equally strong for every individual cluster: M5 provides
only tentative evidence for an intrinsic spread, while NGC~2298 does not by
itself provide a high-confidence confirmation of the trend. We therefore
interpret the metallicity dependence as an ensemble property of the
homogeneous sample rather than as a sequence established independently by
every cluster.

Recent high-precision differential abundance measurements
of M92 provide a particularly useful constraint on this
interpretation. \citet{Kirby2026} find an intrinsic Eu
dispersion of
$\sigma_{\rm Eu}=0.219^{+0.152}_{-0.074}$ dex
among first-population stars, while the corresponding
Fe dispersion is constrained to
$\sigma_{\rm Fe}<0.038$ dex (95\% confidence).
This contrast suggests that rare $r$-process enrichment
can produce substantial abundance variations even in
environments where iron has already become relatively
homogeneous. Such chemical structure can persist through
turbulent transport and may subsequently be inherited
by stars forming from the enriched gas.

To compare this stellar abundance field with the simulations, we adopt
sampling regions with $R=39$ pc, a characteristic scale motivated by the
turbulent structure of the simulated disk
\citep{2025ApJ...994..193B,2026ApJ...997...33C,2026ApJ..1004L...9B}.
For each region, we calculate the median [Eu/Fe] abundance as defined in
Section~\ref{sec:methods}. The crosses in the left panel show the mean of
these cloud medians, while their vertical extent shows the dispersion of the
median abundances from cloud to cloud. The simulation points therefore
characterize both the typical abundance of the sampled regions and the
spatial variation in that characteristic abundance. At low [Fe/H], different
regions can have substantially different median $r$-process abundances,
whereas this cloud-to-cloud variation decreases as chemical evolution and
turbulent redistribution proceed.

The right panel addresses a complementary question: how much abundance
variation remains \emph{within} the sampled regions? For each $R=39$ pc
region we calculate the internal standard deviation of [Eu/Fe],
$\sigma_j(R)$, and average this quantity over the ensemble of regions. This
cloud-scale dispersion provides the more direct pre-collapse comparison with
the observed star-to-star dispersion within an individual globular cluster.
For reference, we also show the $1\sigma$ [Eu/Fe] dispersion of all gas
satisfying the temperature and density selection. The two simulated curves
therefore characterize different levels of the abundance structure: one
measures the typical internal variation within finite regions, while the
other measures the larger dispersion of the selected gas considered as a
whole.

The substantial dispersion of the selected gas as a whole provides important
context for the finite-region result. It demonstrates that the cloud-scale
variations are drawn from a chemically heterogeneous abundance field rather
than being peculiar to the adopted sampling radius. This interpretation is
also supported by Figure~\ref{fig:EuFe_scale}, where comparable spatial
variations are recovered over a broad range of $R$. Changing the sampling
radius changes how much of the underlying abundance structure is averaged
within a region, but the presence of significant chemical heterogeneity is
not restricted to a single choice of scale.

Here [Fe/H] provides a measure of chemical evolution within the simulation,
rather than a universal mapping between metallicity and age. Both the
box-wide and finite-region abundance variations decrease as the gas becomes
chemically enriched and the ejecta are redistributed. The comparison with
the globular clusters is therefore qualitative: the simulations provide
substantial $r$-process variance at low [Fe/H], with less variance available
at higher metallicity, broadly consistent with the behavior suggested by the
cluster ensemble. We do not expect a point-by-point correspondence between
individual clusters and the simulated curves.

Figure~\ref{fig:obs} therefore separates several complementary measures of
the chemical structure. The left panel measures how the characteristic
abundance varies from one finite region to another. The right panel measures
the abundance variation within those regions and compares it with the larger
dispersion of the selected gas as a whole. Figure~\ref{fig:EuFe_scale}
further shows that the region-to-region variation persists across a range of
sampling radii. Together, these measurements show that the simulated ISM
contains a substantial reservoir of $r$-process heterogeneity, part of which
remains present within finite regions of the size relevant to prospective
cluster-forming environments.

The absolute amplitude of this pre-collapse variance should not, however, be
identified directly with the final stellar abundance dispersion. Our
calculations do not resolve the molecular thermochemistry, self-gravitating
cloud assembly, collapse, fragmentation, or star formation of an individual
globular-cluster progenitor. Turbulence plays a central role throughout cloud
assembly and star formation
\citep{2004RvMP...76..125M,2007ARA&A..45..565M}. More directly, calculations
that follow chemical inhomogeneities through turbulent cloud assembly and
collapse show that these processes can substantially reduce the abundance
scatter ultimately inherited by the stars
\citep{2014Natur.513..523F,2018MNRAS.481.5000A}. Stellar winds and other
forms of feedback provide additional sources of energy and momentum once
massive stars begin to form
\citep{2014MNRAS.442.2701R,2020ApJ...899L..30G}.

How much of the pre-collapse chemical structure survives should also depend
on the time available for this redistribution. If cloud assembly and star
formation proceed rapidly compared with turbulent mixing on the relevant
physical scales, more of the initial chemical structure may be preserved.
Conversely, if turbulent redistribution operates for longer before the
stellar abundances are locked in, the resulting stellar population should
inherit a smaller dispersion. We do not model this competition directly
here. It provides an additional reason to interpret the simulated gas
dispersions as the chemical variance available before cluster formation
rather than as direct predictions of the final stellar abundance scatter.

The systematically larger simulated dispersions are therefore not
unexpected. Indeed, the difference between the dispersion of the selected
gas as a whole, the smaller dispersion within finite sampled regions, and
the still smaller stellar dispersions illustrates the sequence of spatial
selection and subsequent mixing that separates the galactic abundance field
from the final stellar fossil record. The more diagnostic comparison is
whether the simulations provide sufficient pre-collapse variance and whether
that variance exhibits the same broad metallicity dependence as the
observations. Rare enrichment produces substantial abundance variance at low
[Fe/H], and the available variance decreases as the gas becomes chemically
enriched and the ejecta are redistributed. The observed stellar
distributions can then be understood as sampling the fraction of this
pre-existing structure that survives cloud assembly and star formation.

The resulting picture connects the cluster fossil record to the chemical
evolution of its natal environment. Rare enrichment initially produces a
strongly inhomogeneous abundance field. Supernova-driven turbulence
redistributes this material, while spatial averaging determines how much of
that structure is apparent within regions of different physical size.
Subsequent cloud assembly and star formation can further reduce the
pre-existing variance before the abundances are locked into stars. The
observed decrease in $r$-process dispersion toward increasing metallicity may
therefore preserve information about the chemical structure and mixing
history of the gas from which the clusters formed.

This interpretation is deliberately agnostic about the astrophysical source
of the $r$-process material. The rare enrichment event may be a rare
supernova or a compact-object merger. The nature of the progenitor determines
when and where the ejecta are deposited, and in particular whether enrichment
occurs within the relevant star-forming environment. Once deposited, however,
the ejecta are redistributed by the turbulent ISM represented in the
simulation. Globular clusters can thus preserve information not only about
rare nucleosynthesis, but also about what happened to its products between
their injection into the ISM and their incorporation into stars.

\section{Summary and Conclusions}
\label{sec:conclusions}

We have used parsec-scale simulations of rare-element transport in a
supernova-driven turbulent disk to ask whether the $r$-process variations
observed in Galactic globular clusters can be inherited from chemical
structure in their natal environments. Rare enrichment produces a highly
inhomogeneous [Eu/Fe] field whose dispersion decreases as chemical evolution
and turbulent redistribution proceed. This structure remains apparent when
the gas is sampled over finite spatial scales: distinct regions can have
different characteristic abundances, while substantial abundance variation
also persists within individual regions.

This provides a natural connection to the globular-cluster abundance record.
Several of the most metal-poor clusters show substantial internal
$r$-process variations, while the homogeneous observational sample suggests
smaller variations toward increasing [Fe/H]. The simulations exhibit the
same broad behavior, providing substantial pre-collapse $r$-process variance
at low metallicity that decreases with chemical evolution. The larger
dispersion of the selected gas as a whole, together with the variations
recovered across a range of sampling radii, demonstrates that this result is
not peculiar to a single adopted cloud scale. The simulated dispersions
remain larger than those observed in stars, as expected if subsequent cloud
assembly, collapse, and star formation further redistribute the material
before stellar abundances are established.

Our results therefore support an interpretation in which at least some of
the $r$-process variations observed in globular clusters are inherited from
their natal galactic environments. This conclusion does not require a
specific $r$-process progenitor: a rare supernova or compact-object merger
can provide the enrichment, provided that its ejecta are deposited into the
relevant star-forming environment. Globular clusters may thus preserve a
fossil record not only of rare nucleosynthesis, but also of the chemical
structure that survived between enrichment of the ISM and the formation of
their stars.

\section*{Acknowledgments}
We are grateful to Anne Noer Kolborg for many insightful discussions and for her key contributions to the analysis and physical interpretation that helped shape this work. We also thank James Beattie for illuminating discussions on turbulent mixing and for helping us interpret the scale-dependent chemical fluctuations seen in the simulations.
This research was also made possible through funding from the Lamat Institute \citep{2025NatAs...9.1770Q} and support by the Heising-Simons Foundation, the Danish National
Research Foundation (DNRF132), and NSF (AST-2447606,  AST-2150255 and AST-1911206). This work made use of an HPC facility funded by a grant from VILLUM FONDEN (project number 16599).   The authors used OpenAI Codex \citep{OpenAI2025_Codex} for assistance with manuscript editing.

\FloatBarrier

\nocite{*}
\bibliography{bibliography}{}
\bibliographystyle{aasjournal}


\end{document}